\documentclass[aps,twocolumn,amsmath,amssymb,preprintnumbers]{revtex4}
\usepackage{amsmath} \usepackage{amsfonts} \usepackage{amssymb}
\usepackage{bbm}
\usepackage{epsfig}
\usepackage{graphics}
\usepackage{graphicx}

\newcommand{\eeq}{\end{equation}}
\newcommand{\beq}{\begin{equation}}
\newcommand{\ba}{\begin{array}}
\newcommand{\ea}{\end{array}}
\newcommand{\bea}{\begin{eqnarray}}
\newcommand{\eea}{\end{eqnarray}}
\newcommand{\vev}[1]{\langle #1\rangle}
\newcommand{\vp}{\varphi}

\newcommand{\preprintno}[1]{\vspace{-2cm}{\normalsize\begin{flushright}#1\end{flushright}}\vspace{1cm}}

\begin{document}

\title{\preprintno{HD-THEP-08-33}Competing bounds on the present-day time variation of fundamental constants}

\author{Thomas Dent, Steffen Stern and Christof Wetterich}

\affiliation{Institut f\"ur Theoretische Physik, Universit\"at Heidelberg\\
Philosophenweg 16, D-69120 Heidelberg}

\begin{abstract}\noindent
We compare the sensitivity of a recent bound on time variation of the fine structure constant from optical clocks with bounds on time varying fundamental constants from atomic clocks sensitive to the electron-to-proton mass ratio, from radioactive decay rates in meteorites, and from the Oklo natural reactor. Tests of the Weak Equivalence Principle (WEP) also lead to comparable bounds on present variations of constants. The ``winner in sensitivity'' depends on what relations exist between the variations of different couplings in the standard model of particle physics, which may arise from the unification of gauge interactions. WEP tests are currently the most sensitive within unified scenarios. A detection of time variation in atomic clocks would favour dynamical dark energy and put strong constraints on the dynamics of a cosmological scalar field.
\end{abstract}

\maketitle

Recently a stringent limit on a possible time variation of the fine structure constant $\alpha$ has been obtained from direct comparison of optical frequencies in an Al/Hg ion clock \cite{Rosenband}:
\beq \label{eq:alphabound}
 d \ln \alpha / dt = (-1.6 \pm 2.3) \times 10^{-17} y^{-1} .
\eeq
Furthermore, by comparing different atomic transitions over years in the laboratory one can put stringent bounds on the time variation of the electron-proton mass ratio $\mu \equiv m_e / m_p$ \cite{Peik:2006}. A recent evaluation \cite{Blatt:2008} of atomic clock data gives
\beq \label{eq:mubound}
  d \ln \mu / dt = (1.5 \pm 1.7) \times 10^{-15} y^{-1} .
\eeq

Within quantum field theory such time variations can be associated to the time evolution of some scalar field \cite{Jordan3739,Bekenstein:1982eu}. They are a typical feature of quintessence cosmologies \cite{Wetterich88}, where a dynamical dark energy arises from a time-varying scalar field. Its coupling to atoms or photons results in ``varying fundamental constants'' \cite{Wetterich:1987fk,DvaliZ}. The connection between cosmologically evolving scalar fields and varying couplings has been explored in various theoretical frameworks \cite{DamourP,Sandvik:2001rv,Gasperini:2001pc,Olive:2001vz,Uzanreview,Parkinson:2003kf}. 

There are also bounds on possible variations over cosmological periods, ranging from nucleosynthesis \cite{DSW07} to observations of molecular and atomic spectra at high redshift \cite{Kanekar,Murphy:2003mi}. The most restrictive such astrophysical bound arises from observations of the ammonia inversion spectrum at $z=0.68$ \cite{MurphyNH3}. However, to compare such bounds with present-day experimental limits would require information about the time evolution of the scalar field, and its couplings to atoms and photons, over a substantial range in redshift \cite{Part2}. Thus extrapolations to the present-day variation are strongly model dependent.

However, for more recent ``historical bounds'', concerning nuclear decay rates in meteorites \cite{Olive:2003sq}
and isotopic abundances in the Oklo natural reactor \cite{Damour:1996zw,Gould,Petrov:2005}
the time span to the present is short on cosmological scales. Thus, a linear interpolation in time is meaningful in most theoretical models. By modelling nuclear reactions occurring in the Oklo mine about $1.8\times 10^9$ years ago, one can bound the variation of $\alpha$ over this period. For a linear time dependence, one derives a conservative bound \cite{Petrov:2005}
\beq
\label{eq:OkloBound}
 | \dot{\alpha} / \alpha | \leq 3 \times 10^{-17} y^{-1}.
\eeq
Note that this result concerns varying $\alpha$ only. If other parameters affecting nuclear forces, in particular light quark masses, are allowed to vary, we incorporate the resulting uncertainty by multiplying the bound \eqref{eq:OkloBound} by a factor three. However this procedure, which assumes the absence of ``fine tuned'' cancellations of effects from variations of various couplings, is somewhat arbitrary. 

The change of decay rates of long-lived $\alpha$- and $\beta$-decay isotopes in meteorites may be sensitive probes for a recent (over the last 4 billion years) variation of constants. The best bound concerns $\beta$-decay of $^{187}$Re in meteorites which formed about $4.6 \times 10^9$ years ago. With a linear interpolation the fractional rate of change of the decay rate $\lambda_{187}$ is bounded by \cite{Olive:2003sq,FujiiMeteorites}
\beq \label{eq:metbound}
  d \ln {\lambda_{187}}/d t = (-7.2 \pm 6.9) \times10^{-12}\,{\rm y}^{-1},
\eeq
and we find for the decay rate dependence on fundamental parameters \cite{Part1}
\begin{multline}
  \Delta \ln (\lambda_{187}/m_N) \simeq 
  -2.2\times 10^4 \Delta \ln \alpha - 1.9\times 10^4 \Delta \ln (\hat{m}/m_N) \\ 
  + 2300\, \Delta \ln (\delta_q/m_N) -580\, \Delta \ln (m_e/m_N).
\end{multline}
Here $m_N$ is the average nucleon mass $m_N = (m_p + m_n)/2$, $\hat{m}$ the average light quark mass $\hat{m} = (m_u + m_d)/2$, and $\delta_q = m_d - m_u$ the light quark mass difference.

In this Letter, we compare the sensitivity of the bounds \eqref{eq:alphabound}, \eqref{eq:mubound}, \eqref{eq:OkloBound}, \eqref{eq:metbound}, and also compare these with the sensitivity of bounds on the violation of the Weak Equivalence Principle (WEP). Such comparisons involve assumptions about relations between the variations of different couplings of the standard model of particle physics. The comparison with WEP bounds will also use cosmological constraints on the time evolution of a scalar field.

\paragraph*{Unification of gauge interactions.} 
A comparison between variations of different couplings becomes possible given certain relations resulting from a theory of grand unification (GUT) \cite{CalmetFritzsch, Part1}, with a unified coupling constant $\alpha_X$ and GUT breaking scale $M_X$. The unification of the electromagnetic and strong gauge couplings relates a variation of $m_N / M_X$ to a variation of $\alpha$. The other crucial parameter is the variation of the Fermi scale given by the vacuum expectation value of the Higgs doublet $\vev{\phi}$. Different assumptions for this variation define different scenarios for coupling variations in GUTs. We consider variations in the set of ``fundamental'' parameters
\beq
G_k = \{ G_{\rm N}, \alpha, \vev{\phi}, m_e, \delta_q, \hat{m} \},
\eeq
where $G_{\rm N}$ is Newton's constant. All couplings with non-zero mass dimensions are measured in units of the strong interaction scale $\Lambda_{QCD}$ which we may keep fixed. To a good approximation we can replace $\Lambda_{QCD}$ by $m_N$, such that a relative variation $\Delta \ln m_e$ denotes a variation of the mass ratio $\Delta \ln (m_e / m_N)$.

We consider the thesis that relative variations of the parameters $G_k$ are proportional to one nontrivial variation with fixed constants of proportionality. Here, we are concerned only with the evolution over a recent cosmological period $z \lesssim 0.5$. If the variation of the unified gauge coupling $\Delta \ln \alpha_X$ (away from its present value) is nonvanishing, we may write
\beq \label{dkdef}
  \Delta \ln G_k = d_k \Delta \ln \alpha_X
\eeq
for some constants $d_k$ which depend on the specific GUT model. If $\alpha_X$ does not vary, we consider instead $\Delta \ln \vev{\phi}/M_X$ on the RHS of this equation. We will assume here constant Yukawa couplings, i.e.
\beq
 \Delta \ln \frac{m_e}{M_X} = \Delta \ln \frac{\delta_q}{M_X} = \Delta \ln \frac{\hat{m}}{M_X} = \Delta \ln \frac{\vev{\phi}}{M_X}.
\eeq
This leaves three independent ``unification coefficients'' whose choice will define the grand unified scenarios: 
\bea \label{eq:unifdefs}
 \Delta \ln \frac{M_X}{M_{\rm P}} = d_M l, \quad \Delta \ln \alpha_X = d_X l, \nonumber \\
 \Delta \ln \frac{\vev{\phi}}{M_X} = d_H l, \quad \Delta \ln \frac{m_S}{M_X} = d_S l.
\eea
Here $m_S$ is a typical mass scale for supersymmetric partners, for scenarios with supersymmetry. Depending on the GUT scenario, either $l = \Delta \ln \alpha_X$ ($d_X = 1$) or $l = \Delta \ln \vev{\phi}/M_X$ ($d_H = 1$). The different scenarios investigated here are displayed in Table~\ref{tab:Scenarios}. 
\begin{table}[htb]
\center
\begin{tabular}{|lc|ccccc|}
\hline
Scenario & $l$              & $d_M$ & $d_X$  & $d_H$ & $d_S$ & $\alpha_X$ \\
\hline
2                        & $\alpha_X$       & 0& 1& 0& 0& 1/40\\
2S                       & $\alpha_X$       & 0& 1& 0& 0& 1/24\\
3                        & $\vev{\phi}/M_X$ & 0& 0& 1& 0& 1/40\\
4                        & $\vev{\phi}/M_X$ & 0& 0& 1& 1& 1/24\\
5, $\tilde{\gamma} = 42$ & $\alpha_X$       & 0& 1& 42& 0& 1/40\\
6, $\tilde{\gamma} = 70$ & $\alpha_X$       & 0& 1& 70& 70& 1/24\\
6, $\tilde{\gamma} = 25$ & $\alpha_X$       & 0& 1& 25& 25& 1/24\\
\hline
\end{tabular}
\caption{Unified scenarios}
\label{tab:Scenarios}
\end{table}
The parameter $\tilde{\gamma} \equiv d_H / d_X$ describes variation of $\vev{\phi} / M_X$, which is essentially a free parameter, given theoretical uncertainties in any relations between the Fermi scale and the unification scale. Scenarios 2S, 4 and 6 contain  superpartners. We also will consider a (non-unified) scenario where only $\alpha$ varies. For each scenario the variations of all fundamental parameters $G_k$ can be expressed in terms of one varying GUT parameter (either $\Delta \ln \alpha_X$ or $\Delta \ln \vev{\phi}/M_X$), via the coefficients $d_k$ derived in \cite{Part1}, thus variations of different parameters (at similar times or redshifts) may be compared.

In Table~\ref{tab:PresentVariationErrors} we compare the precision of bounds on fractional variations of the fundamental parameter in each scenario ($\alpha$, $\alpha_X$ or $\vev{\phi}/M_X$). 
\begin{table}
\center
\begin{tabular}{|lc|ccccc|}
\hline
 & & \multicolumn{5}{c|}{ Error on $d \ln X/dt$ ($10^{-15}y^{-1}$)} \\
\hline
Scenario & $X$              & Al/Hg & Clocks ($\mu$) & Oklo & ${^{187}}$Re & WEP \\
\hline
$\alpha$ only & $\alpha$    &  0.023 & -      & 0.033 & 0.32  & 6.2    \\
2        & $\alpha_X$       & 0.027  & 0.074  & 0.12  & 0.015 & 0.007 \\
2S       & $\alpha_X$       & 0.044  & 0.12   & 0.19  & 0.026 & 0.012 \\
3        & $\vev{\phi}/M_X$ & 12.4   & 2.6    & 54    & 0.53  & 0.33   \\
4        & $\vev{\phi}/M_X$ & 1.78   & 6.2    & 7.7   & 1.2   & 0.35   \\
5, $\tilde{\gamma} = 42$ & $\alpha_X$ & 0.024  & 0.42  & 0.11  & 0.069 & 0.013  \\
6, $\tilde{\gamma} = 70$ & $\alpha_X$ & 0.016  & 0.30  & 0.070 & 0.049 & 0.008 \\
6, $\tilde{\gamma} = 25$ & $\alpha_X$ & 0.027  & 0.25  & 0.12  & 0.056 & 0.011 \\
\hline
\end{tabular}
\caption{Competing bounds on present time variations. For each scenario we give the $1\sigma$ uncertainties of null bounds on $d(\ln X)/dt$ in units $10^{-15}{\rm y}^{-1}$. The column ``Clocks'' results from atomic clock bounds on $\mu$, the recent Al/Hg limit \cite{Rosenband} on $\alpha$ variation being treated separately. The Oklo bound on $\alpha$ variation is rescaled as discussed in the text.}
\label{tab:PresentVariationErrors}
\end{table}
The lowest number indicates the most sensitive bound. Considering Al/Hg, other clocks, Oklo and the ${^{187}}$Re decay from meteorites, if only $\alpha$ varies, atomic clock experiments are already the most sensitive, surpassing the previous best bound from the Oklo reactor. If only the unified coupling $\alpha_X$ varies, the meteorite bound is still somewhat stronger than atomic clocks. The same holds if only the ratio $\vev{\phi} / M_X$ varies. For combined variations of $\alpha_X$ and $\vev{\phi}/M_X$ in scenarios 5 and 6, the ``sensitivity winner'' is again the laboratory bound. Thus, laboratory experiments have now reached the sensitivity of the ``historical'' bounds; a modest further increase in sensitivity, especially for variations of $\mu$, will make them the best probes in all scenarios.

\paragraph*{Quintessence and violation of the WEP.} 
Table \ref{tab:PresentVariationErrors} shows a further competitor for the ``sensitivity race'': bounds on possible violations of the Weak Equivalence Principle. In quantum field theory, any time variation of couplings must be associated to the time evolution of a field. This field may describe a new ``fundamental particle'', or stand for the expectation value of some composite operator. We consider the simplest hypothesis, that the field is a scalar, such that its time-varying expectation value preserves rotation and translation symmetry locally, as well as all gauge symmetries of the standard model. We may identify this scalar field with the ``cosmon'' field $\vp$ of dynamical dark energy or quintessence models. In these models the cosmon field contributes a homogeneous and isotropic energy density in the Universe, leading to dynamical dark energy \cite{Wetterich88}. A time variation of fundamental couplings is a generic prediction of such models \cite{Wetterich:1987fk}. Alternatively the scalar could also be cosmologically insignificant because it contributes too little to the energy density \cite{DamourP}. We treat both scenarios within the same framework of a dynamical cosmon. They are then only distinguished by a different fraction of the cosmological energy density $\Omega_h$ attributed to the scalar field.

The cosmon couples to the Standard Model particles or fields. Typically its coupling is not precisely proportional to the mass of test particles, and therefore the cosmon mediates a ``fifth force'' violating the universality of free fall. 
Two test bodies $b,c$ with 
different composition will experience different accelerations towards a common source, due to their generally different ``cosmon charge'' per mass. The deviation from the universality of free fall is given by the differential acceleration or E{\" o}tv{\" o}s parameter $\eta$:
\beq \label{etadef}
 \eta^{b-c} \equiv 2\frac{a_b-a_c}{a_b+a_c},
\eeq
where $a_{b,\,c}$ are the accelerations of the two test masses towards the source. 
The cosmon couplings to atoms and photons determine both the outcome of tests of the WEP, 
and the time variation of ``constants'' in recent cosmological epochs including the present. 
This is the basic reason why WEP tests and bounds on time-varying couplings variations can be compared for sensitivity. For this comparison 
one needs additional information on the rate of change of the expectation value of the scalar field, since an observable time variation involves both the cosmon coupling and its time derivative. The latter can be expressed in terms of cosmological observables, namely the fraction in dark energy contributed by the cosmon, $\Omega_h$, and its equation of state, $w_h$, as $\dot{\vp}^2 / 2 = \Omega_h (1 + w_h) \rho_c$. Here $\rho_c = 3 H^2 M_P^2$ is the critical energy density of the Universe.

Thus the differential acceleration $\eta$ 
can be related to the present time variation of couplings and to cosmological parameters. Taking the fine structure constant as varying parameter, we find  \cite{Wetterich:2002ic, Dent06, Part2} at the present epoch
\beq
\label{eq:alphacosmoeta}
 \left| \frac{\dot{\alpha}/\alpha}{10^{-15}{\rm y}^{-1}} \right| = \left|\frac{\Omega_h (1 + w_h)}{F} \right|^{1/2} \left| \frac{\eta}{3.8 \times 10^{-12}} \right|^{1/2} .
\eeq
The ``unification factor'' $F$ encodes the dependence on the precise unification scenario (defined in Tab.~\ref{tab:Scenarios}) and on the composition of the test bodies. The values of $F$ were computed in \cite{Part2} for the Be/Ti test masses used in the best current test of WEP \cite{Schlamminger:2007ht} and are shown in 
\begin{table}
\center
\begin{tabular}{|c||c|c|c|c|c|c|c|}
\hline
Scenario & $\alpha$ only & 2 & 3 & 4 & 5, $\tilde{\gamma}\! =\! 42$ & 6, $\tilde{\gamma}\! =\! 70$ & 6, $\tilde{\gamma}\! =\! 25$ \\ 
\hline 
$F$ (Be-Ti)& $-9.3\, 10^{-5}$ & 95  & -9000 & -165  & -25  & -26  & 41 \\
\hline
\end{tabular}
\caption{Values of $F$ for a WEP experiment using Be-Ti masses, in different unified scenarios.} 
\label{tab:F}
\end{table}
Table~\ref{tab:F}. 

Once $F$ is fixed, the relation \eqref{eq:alphacosmoeta} allows for a direct comparison between the sensitivity of measurements of $\eta$ with measurements of $\dot{\alpha}/\alpha$ from laboratory experiments, or bounds from recent cosmological history, such as from the Oklo natural reactor or the isotopic composition of meteorites, provided we can use cosmological information for $\Omega_h (1 + w_h)$. The most stringent test of WEP gives 
\beq \label{Schlameta}
 \eta = (0.3\pm 1.8)\times 10^{-13}
\eeq
for test bodies of Be and Ti composition \cite{Schlamminger:2007ht}, with the gravitational source taken to be the Earth. We will assume initially that the time varying field is responsible for the dark energy in the Universe: then the cosmological observations imply $\Omega_h \simeq 0.73$ and $w_h\leq -0.9$.

The resulting WEP bounds on the time variation of $\alpha$, or of the appropriate unified coupling, are displayed in the last column of Tab.~\ref{tab:PresentVariationErrors}. For all unification scenarios these bounds are the most severe. At present, clock experiments 
win the ``sensitivity race'' only if $\alpha$ is the only time-varying coupling. Varying only $\alpha$ is not a very plausible scenario in the context of electroweak unification and perhaps grand unification. 

WEP bounds become even more restrictive if the scalar field plays no significant role in cosmology. In this case $\Omega_h$ is small, say $\Omega_h < 0.01$. Even for substantially larger $1 + w_h$, say $w_h = 0$, the product $\Omega_h (1+ w_h)$ will not exceed the maximal value for the cosmon, $\Omega_h (1 + w_h) \lesssim 0.07$. The relevant limit is actually an observational bound on the scalar kinetic energy, and therefore on $|\dot{\vp}|$, independent of the precise scalar model and assumptions on its overall role within cosmology. More severe restrictions on $\Omega_h (1+w_h)$ -- usually expressed as limits on $w_h$ -- will further enhance the sensitivity of the WEP bound for time varying couplings.

The converse argument is equally strong: if a time variation of the fine structure constant is observed close to the present limit, most unification scenarios can be ruled out. Only scenarios with a value $|F| \lesssim 10$ would remain compatible with the WEP bounds. In view of the different scenarios investigated here this would require a certain tuning of parameters $d_k$ in Eq.~\eqref{dkdef}
to achieve a small differential coupling of the cosmon in the WEP test with Be/Ti masses. Further precise tests with materials of different composition will make it even more difficult to ``hide'' the cosmon coupling to photons needed for $\dot{\alpha}/\alpha \neq 0$ from detection in the WEP tests. Any detection of a time variation in $\mu$ in atomic clocks close to the current best sensitivity would be even harder to reconcile with WEP tests. In this case the varying electron and/or proton masses are due to the cosmon couplings to electrons or protons, which are directly probed in the WEP experiments.

A detection in atomic clock experiments of a nonzero coupling variation near the present experimental bounds would put important constraints on cosmology. Firstly, it would require some field to be evolving in the present cosmological epoch. No such candidate field is present in standard $\Lambda$CDM cosmology where a cosmological constant determines dark energy. Clearly, a time variation of couplings would rule out the minimal $\Lambda$CDM model. Dynamical dark energy models with a time-varying field that couples to photons and atoms would be favoured. Furthermore, a combination of time-varying couplings with present WEP bounds on $\eta$ would provide a {\em lower bound} on the kinetic energy of this field. In case of a scalar field we find the bound 
\beq
\label{eq:BoundOnOmegaw}
 \Omega_h (1+w_h) \gtrsim 3.8 \times 10^{18} F (\partial_t \ln \alpha)^2 
 \eta_{\rm max}^{-1} \, .
\eeq
Here $\partial_t \ln \alpha$ is given in units of y$^{-1}$, and $\eta_{\rm max} \simeq 1.8 \times 10^{-13}$ is the current experimental limit on the differential acceleration of test bodies of different composition. Thus, if $|\partial_t \ln \alpha |$ is nonzero and not too small, $w_h$ {\em cannot}\/ be arbitrarily close to $-1$ (a cosmological constant); nor can the contribution of the scalar to the dark energy density be insignificant. The precise bound depends on the ``unification factor'' $F$, which depends on the different scenarios of unification and also depends on the composition of the experimental test bodies. For a reasonable lower bound $|F| > 10$ a value $|\dot{\alpha}/\alpha| = 10^{-17} y^{-1}$ would imply $\Omega_h (1+w_h) > 0.02$, yielding $w_h > -0.97$ for $\Omega_h = 0.73$. Clearly, there would be a competition between laboratory (atomic clock plus WEP) lower bounds and cosmological upper bounds on $w_h$.

The race for the best bounds on the present time variation of fundamental couplings is open, independently of other possible interesting observations of varying couplings in early cosmological epochs. The precision of WEP-bounds on $\eta$ and atomic clock bounds on $\dot{\alpha}/\alpha$ and $\dot{\mu}/\mu$ is expected to increase. Laboratory experiments will then be clearly more sensitive than  ``historical bounds'' from meteorites or the Oklo natural reactor. At present the WEP bounds are leading in the race; the future will show which experimental approach is more sensitive. The largest restrictive power or the largest discovery potential arises from a combined increase in sensitivity of both WEP and clock experiments. Their mutual consistency, as well as consistency with cosmological bounds, places severe restrictions on the underlying models in case of a positive signal.



\begin{thebibliography}{99} 


\bibitem{Rosenband}
  T.~Rosenband {\it et al.}, 
  Science {\bf 319} (2008), 1808.

\bibitem{Peik:2006}
  E.~Peik, B.~Lipphardt, H.~Schnatz, C.~Tamm, S.~Weyers and R.~Wynands,
  arXiv:physics/0611088.

\bibitem{Blatt:2008}
  S.~Blatt {\it et al.},
  Phys.\ Rev.\ Lett.\  {\bf 100} (2008) 140801.

\bibitem{Jordan3739}
   P.~Jordan,
   Naturwiss. {\bf25} 513 (1937);
   Z. Physik {\bf113} 660 (1939).

\bibitem{Bekenstein:1982eu}
  J.~D.~Bekenstein,
  Phys.\ Rev.\  D {\bf 25} (1982) 1527.

\bibitem{Wetterich88}
  C.~Wetterich,
  Nucl.\ Phys.\  B {\bf 302} (1988) 668.

\bibitem{Wetterich:1987fk}
  C.~Wetterich,
  Nucl.\ Phys.\  B {\bf 302} (1988) 645.

\bibitem{DvaliZ}
  G.~R.~Dvali and M.~Zaldarriaga,
  Phys.\ Rev.\ Lett.\  {\bf 88} (2002) 091303.

\bibitem{DamourP}
  T.~Damour and A.\,M.~Polyakov,
  Nucl.\ Phys.\ B {\bf 423} (1994) 532.

\bibitem{Sandvik:2001rv}
  H.~B.~Sandvik, J.~D.~Barrow and J.~Magueijo,
  Phys.\ Rev.\ Lett.\  {\bf 88} (2002) 031302.

\bibitem{Gasperini:2001pc}
  M.~Gasperini, F.~Piazza and G.~Veneziano,
  Phys.\ Rev.\  D {\bf 65} (2002) 023508.

\bibitem{Olive:2001vz}
  K.~A.~Olive and M.~Pospelov,
  Phys.\ Rev.\  D {\bf 65} (2002) 085044;
%
  E.~J.~Copeland, N.~J.~Nunes and M.~Pospelov,
  Phys.\ Rev.\  D {\bf 69} (2004) 023501.

\bibitem{Uzanreview}
  J.-P.~Uzan,
  Rev.\ Mod.\ Phys.\  {\bf 75} (2003) 403.

\bibitem{Parkinson:2003kf}
  D.~Parkinson, B.~A.~Bassett and J.~D.~Barrow,
  Phys.\ Lett.\  B {\bf 578} (2004) 235;
%
  N.~J.~Nunes and J.~E.~Lidsey,
  Phys.\ Rev.\  D {\bf 69} (2004) 123511.

\bibitem{DSW07}
  T.~Dent, S.~Stern and C.~Wetterich,
  Phys.\ Rev.\  D {\bf 76} (2007) 063513;
  A.~Coc {\em et al.}, 
  Phys.\ Rev.\  D {\bf 76} (2007) 023511.

\bibitem{Kanekar}
  N.~Kanekar,
  Mod.\ Phys.\ Lett.\  A {\bf 23} (2008) 2711;
%
  V.~V.~Flambaum,
  Eur.\ Phys.\ J.\ ST {\bf 163} (2008) 159.

\bibitem{Murphy:2003mi}
  M.\,T.~Murphy {\it et al.}, 
  Lect.\ Notes Phys.\  {\bf 648}, 131 (2004).

\bibitem{MurphyNH3}
  M.\,T.~Murphy, V.\,V.~Flambaum, S.~Muller and C.~Henkel,
  Science {\bf 320}, 1611 (2008)
  [0806.3081 [astro-ph]].
  
\bibitem{Part2}
  T.~Dent, S.~Stern and C.~Wetterich,
  arXiv:0809.4628.
  
\bibitem{Olive:2003sq}
  K.~A.~Olive {\em et al.}, 
  Phys.\ Rev.\  D {\bf 69}, 027701 (2004)

\bibitem{Damour:1996zw}
  T.~Damour and F.~Dyson,
  Nucl.\ Phys.\  B {\bf 480} (1996) 37;
  Y.~Fujii {\it et al.},
  Nucl.\ Phys.\  B {\bf 573} (2000) 377.

\bibitem{Gould}
  C.\,R.~Gould, E.\,I.~Sharapov and S.\,K.~Lamoreaux,
  Phys.\ Rev.\  C {\bf 74} (2006) 024607.

\bibitem{Petrov:2005}
  Yu.\,V.~Petrov {\it et al.}, 
  Phys.\ Rev.\  C {\bf 74} (2006) 064610. 

\bibitem{FujiiMeteorites}
  Y.~Fujii and A.~Iwamoto,
  Phys.\ Rev.\ Lett.\  {\bf 91} (2003) 261101.

\bibitem{Part1}
  T.~Dent, S.~Stern and C.~Wetterich,
  Phys.\ Rev.\ D {\bf 78} (2008) 103518.

\bibitem{CalmetFritzsch}
  X.~Calmet and H.~Fritzsch,
  Eur.\ Phys.\ J.\  C {\bf 24} (2002) 639;
%
  P.~Langacker, G.~Segre and M.~J.~Strassler,
  Phys.\ Lett.\  B {\bf 528} (2002) 121.

\bibitem{Wetterich:2002ic}
  C.~Wetterich,
  JCAP {\bf 0310} (2003) 002.

\bibitem{Dent06}
  T.~Dent,
  JCAP {\bf 0701} (2007) 013.


\bibitem{Schlamminger:2007ht}
  S.~Schlamminger {\it et al.}, 
  Phys.\ Rev.\ Lett.\  {\bf 100} (2008) 041101.






\end{thebibliography}
\end{document}